\documentclass{aa}
\usepackage{graphicx}
\begin{document}
   \title{Radio Luminosities and Classificatory Criteria of BL Lacertae Objects}

   \author{D. C. Mei
          \thanks{email:kmdcmei@public.km.yn.cn}
          \and
          L.Zhang
          \and
          Z. J. Jiang
           \and
          B.Z.Dai}

   \offprints{D.C.Mei}

   \institute{Department of Physics,
Yunnan University, Kunming 650091, P. R. China}

   \date{Received ; accepted }
   \authorrunning{D.C.Mei et al.}
   \titlerunning{Radio Luminosities and Classificatory Criteria of BL Lacertae Objects }

   \abstract{
  Using the sample of radio selected BL Lacertae objects (RBLs) and X-ray selected
BL Lacertae objects (XBLs)  presented by Sambruna et al. (1996),
we calculated the luminosities of radio, optical and X-ray of each
source and made the statistical analysis among the luminosities at
different wave-bands, broad-band spectral indices from radio to
X-ray ($\alpha_{\rm rx}$) and peak frequencies ($\nu_p$). Our
results are as follows: (i) there is a positive correlation
between radio luminosity $L_{\rm r}$ and $\alpha_{\rm rx}$ and a
negative correlation between $L_{\rm r}$ and $\nu_p$. High-energy
peak BL Lacs (HBLs) and low-energy peak BL Lacs (LBLs)  can be
distinguished very well, the dividing lines are probably those of
$\log {L_{\rm r}}=43.25$ (erg/sec) and $\alpha_{\rm rx}>$(or $\leq
$)$0.75$ for $L_{\rm r}$ - $\alpha_{\rm rx}$ plot and those of
$\log {L_{\rm r}}\leq 43.25$ (erg/sec) and $\log {\nu_p}>14.7$ for
the $L_{\rm r}$ - $\nu_p$ plot; (ii) there is a weak positive
correlation between optical luminosity $L_o$ and $\alpha_{\rm rx}$
and a  negatively weak correlation between $L_{\rm o}$ and
$\nu_p$; (iii) there is no correlation between X-ray luminosity
$L_X$ and $\alpha_{\rm rx}$ or between $L_X$ and $\nu_p$. From our
analysis, we find that synchrotron radiation is the main X-ray
radiation mechanism  for HBLs while inverse Compton scattering for
LBLs. 
   }
 \maketitle
%

Most BL Lac objects have been identified in either radio survey (
radio-selected BL Lac objects (RBLs) or X-ray survey
(X-ray-selected BL Lac objects (XBLs)). These two kinds of BL Lac
objects exist many significant differences( for example see,
\cite{Le85}; \cite{Gh86}; \cite{St91}). The main differences
between RBLs and XBLs are as follows. First, there is a difference
in spectral energy distributions (SEDs), RBL--like objects show
$\alpha_{\rm rx}>0.75$, while XBL-like objects show $\alpha_{\rm
rx}\leq 0.75$ (see, e.g., \cite{Le85}; \cite{St85}; \cite{Gi90};
\cite{St91}; \cite{Sc93}; Giommi \& Padovani 1994). It should be
pointed out that Padovani \& Giommi (1995) introduced the
distinction between high-energy peak BL Lacs (HBLs) and low-energy
peak BL Lacs (LBLs), for objects which emit most of their
synchrotron power at high (UV-soft-X) and low (far-IR, optical)
frequencies respectively. However, a quantitative distinction can
be drawn on the basis of the ratio between radio and X-ray fluxes.
It can also use the broad-band spectral index $\alpha_{\rm rx}$ to
distinguish HBLs and LBLs: $\alpha_{\rm rx}\leq 0.75$ for HBLs and
$\alpha_{\rm rx}>0.75$ for LBLs (Giommi \& Padovani 1995).
Secondly, there is a difference in the peak frequency
distributions(\cite{Gi95}; \cite{Pa95}; \cite{Ur95}; \cite{Pa96};
\cite{La96}). Sambruna et al.(1996) made a parabolic fit to
multi-band flux data to obtain the peak frequency of the
synchrotron emission, $\nu _{p}$, of sources, when studying
quantitatively how the peak-emission frequency of the synchrotron
emission can be used to distinguish RBL--like and XBL--like
objects. Making use of the peak frequency, $\nu _{p}$, of the
synchrotron emission estimated by Sambruna et al.(1996). Qin et
al.(1999) found that, in the four different regions divided by the
$a_{\textrm{rx}}=0.75$ line and the $\log {\nu
_{\textrm{p}}}=14.7$ line, all RBL--like objects lie in the upper
left region, while most XBL--like objects are within the lower
right region, with only a few sources being located in the lower
left region and no sources being located in the upper right
region. Considering the proper physical origin of the X-ray
emission for different classes of BL Lac objects, we calculated
the peak frequency $\nu_{p}$ of the synchrotron emission of each
source in sample of Sambruna et al. (1996).  It was found that all
RBL-like objects, defined by $\alpha_{\rm rx}>0.75$, are located
in the upper-left region, while all  XBL-like objects, defined by
$\alpha_{\rm rx}\leq 0.75$, are inside the lower-right region. No
sources are in the lower left and upper-right regions, suggesting
that the classificatory criteria in terms of the peak frequency
should be $\log{\nu_{{p}}}=14.7$ and the classificatory criteria
of the $\alpha_{\rm rx}$ is equal to the classificatory criteria
of the $\log{\nu_{{p}}}$ (\cite{Don02}). This provides evidence
supporting what Giommi et al. (1995) proposed: RBL-like and
XBL-like objects can be distinguished by the difference in the
peak frequency of the synchrotron emission.

Sambruna et al.(1996) found that the position of the peak
frequency $\log{\nu_{p}}$ is linked to the luminosity. For more
luminous objects, the peak of the synchrotron power locate at
lower frequencies. Fossati et al.(1998) studied the spectral
energy distributions of three subclass samples of blazars. They
also found that despite the differences in continuum shapes for
sub-classes of blazars, a unified scheme is possible, blazar
continua can be described by luminosity as a fundamental
parameter.

In this paper, we will study how the luminosity of  each energy
band can be used to distinguish RBLs and XBLs. We calculated
luminosity of each energy band for  RBLs and XBLs presented in
Table 1 of Sambruna et al.(1996) in section 2. We present our
analysis results in section 3, and give our discussion and
conclusions in section 4.

\section{ Luminosity of BL Lac objects}

We calculated the luminosities for the sample of RBLs and XBLs
presented by Sambruna et al.(1996). We list the names, red-shifts,
broad-band spectral indices, the energy fluxes at radio, optical
and X-ray bands and peak frequencies of 23 XBLs in Table 1 and 29
RBLs in Table 2, respectively. Using the logrithmic parabolic form
given by Laudau et al (1986) and the data of the energy fluxes at
radio, optical and X-ray bands, Dong et al. (2002) calculated the
peak frequencies of 23 XBLs in Table 1 which are used here.
Sambruna et al.(1996) made a parabolic fit to the radio,
millimeter, IR, optical, and X-ray fluxes data to obtain the peak
frequency, $\nu _{p}$, of the spectral energy, where, for some
XBLs, X--Ray data were not included in the fit. The data of peak
frequency $\nu _{p}$, of the synchrotron emission in Table 2 are
made use of Sambruna et al.(1996) except for objects 1652+398 and
2005-489, because the both sources are HBLs ($\alpha_{\rm rx} <
0.75$). We use the $\nu _{p}$ data  of two sources calculated by
Dong, Mei \& Liang (2002). Therefore, in our sample, there are 27
LBLs which consists of RBLs except for 1652+398 and 2005-489 in
Table 2 and 25 HBLs which consists of 23 XBLs and two RBLs
(1652+398 and 2005-489).

We calculated  rest-frame  luminosities of radio, optical and
X-ray energy band using the fluxes of BL Lac objects in Table 1
and Table 2. For the  K-correction, the flux densities were
multiplied by $(1+z)^{\alpha-1}$, where $\alpha$ is the power-law
spectral index in the appropriate energy band
($F_{\nu}\propto\nu^{-\alpha}$). For RBLs , we used
$\alpha_{r}=0.2,$ and $\alpha_{opt}=1.05$ (\cite{Fa94}), while for
XBLs we used $\alpha_{r}=0$ and $\alpha_{opt}=0.65$(\cite{Fa94}).
We used individual spectral indices in X-ray provided by Sambruna
et al.(1996). The redshifts of 0048-097,1147+245 and 1519-273
presented in table 2 of Sambruna et al.(1996) are not available. A
mean value of $z=0.56$ is accordingly adopted for these sources.
The flux densities were converted to luminosities  using $H_0=75$
Km S$^{-1}$  Mpc$^{-1}$ and $q_0=0.5$, assuming isotropic
emission. We calculated  the radio luminosity $L_{\rm r}$, optical
luminosity $L_{\rm o}$ and X-ray luminosity $L_{\rm x}$ in rest
frame.

\section{Results}

We now consider the relations among radio luminosity, broad-band
spectral index and peak frequency. We plot $L_{\rm r}$ versus
$\alpha_{\rm rx}$ in figure 1a, which shows that the radio
luminosity increases as broad-band spectrum index increases. A
linear regression analysis yields
\begin{equation}
\log{L_{r}}=(6.64\pm0.49)\alpha_{\rm rx}+(38.24\pm 0.37),
\end{equation}
where the correlation coefficient is $r=0.886$ and the chance
probability is $p=2.26\times 10^{-17}$, n=52. Obviously, the
correlation is very strong. From this figure, four different
regions are divided by the lines $\alpha _{\rm{rx}}=0.75$ and
$\log {L_{r}}=43.25$ (${\rm erg s^{-1}} $), all LBLs are located
in the upper-right region; all HBLs are inside the lower-left
region(two RBLs, 1652+398 and 2005-489 are HBLs); no sources are
in the lower-right region and the upper-left region. We plot
$L_{\rm r}$ versus $\nu_p$ in figure 1b, which show  that the
radio luminosity $L_{\rm r}$ decreases as the peak frequency
$\nu_{p}$ increases. The correlation analysis gives
\begin{equation}
\log{L_{\rm r}}=-(0.67\pm0.06)\log{\nu_{p}}+(53.13\pm 0.92),
\end{equation}
where the correlation coefficient is $r =-0.841$ and the chance
probability is $p=4.42\times 10^{-14}$, n=52. The correlation is
also very strong. In the four different regions divided by the
lines $\nu_{{p}}=14.7$ and $\log {L_{r}}=43.25$, all LBLs are
located in the upper-left region; all HBLs are inside the
lower-right region(two RBLs, 1652+398 and 2005-489 are HBLs); no
sources are in the lower-left region and  the upper-right region.

In figure 2, we consider the correlation between the optical
luminosity $L_{\rm o}$ and $\alpha_{\rm rx}$ and that between
$L_{\rm o}$ and $\nu_p$. The results are
\begin{equation}
\log{L_{\rm o}}=(2.45\pm0.45)\alpha_{\rm rx}+(43.94\pm 0.33)
\end{equation}
with $r=0.614$ and $p=2.70\times 10^{-6}$, n=52 and
\begin{equation}
\log{L_{\rm o}}=-(0.28\pm0.05)\log{\nu_{p}}+(49.93\pm 0.68)
\end{equation}
with $r=-0.660$ and $p=2.52\times 10^{-7}$, n=52 respectively. It
can be seen that (i)there are a positive correlation between
$L_{\rm o}$ and $\alpha_{\rm rx}$ and a negative correlation
between $L_{\rm o}$ and $\nu_p$, and (ii)HBLs and LBLs can be
distinguished by the line of $\alpha_{\rm rx}\approx 0.75$ (see
Fig. 2a) or the line of $\nu_p\approx 14.7$ (see Fig. 2b), but
they can not be distinguished by  optical luminosity.

Finally, we study the relation of the X-ray luminosity $L_{\rm x}$
with $\alpha_{\rm rx}$ and $\nu_p$. The correlation analysis reads
\begin{equation}
\log{L_{\rm x}}=-(1.41\pm0.43)\alpha_{\rm rx}+(46.08\pm 0.32)
\end{equation}
for $L_{\rm x}$ versus $\alpha_{\rm rx}$, where $r=-0.42$ and
$p=0.002$.
\begin{equation}
\log{L_X}=(0.11\pm0.05)\log{\nu_{p}}+(43.43\pm 0.73)
\end{equation}
for $L_{\rm x}$ versus $\nu_p$, where $r=0.30$ and $p=0.03$.
Obviously, there is no any correlation between $L_{\rm x}$ and
$\alpha_{\rm rx}$ as well as between $L_{\rm x}$ and $\nu_p$. We
plot the results in Fig. 3. The distributions of HBLs and LBLs in
plot of $L_{\rm x}$ versus $\alpha _{{rx}}$ ( Fig. 3a) and $L_{\rm
x}$ versus $\nu_p$ (Fig. 3b) are confusion although the lines of
$\alpha_x\approx 0.75$ and $\nu_p\approx 14.7$ can be drawn.

 The results of the regression analysis are listed in Table 3.

\section{Discussions and conclusions}

Sambruna et al.(1996) have made the statistical analysis of the
relations of bolometric luminosity  $L_B$ (as estimated from
parabolic fits to the multi-wavelength from radio to X-ray bands)
with position of the peak frequency $\log{\nu_{p}}$ and with the
broad-band spectral index $\alpha_{\rm rx}$ for the blazars. They
found that (i)there is a correlation between $L_B$ and
$\alpha_{\rm rx}$ at $>99$\% confidence, which may indicate that
more luminous objects have steeper $\alpha_{\rm rx}$, and (ii)
more luminous sources have smaller peak frequencies, since the
average bolometric luminosity increases from XBLs to RBLs to flat
spectrum radio quasars (FSRQs), on average, FSRQs are more
luminous and have lower peak frequencies while XBLs are less
luminous and have higher peak frequencies. In their analysis, the
XBLs and RBLs are not distinguished very well in either
$L_B$-$\alpha_{\rm rx}$ plot or $L_B$-$\nu_p$ plot. In this paper,
we have studied the relations of radio, optical and X-ray
luminosities with broad-band spectral index ($\alpha_{\rm rx}$)
and peak frequency ($\nu_p$) for 52 BL Lac objects (including 23
XBLs and 29 RBLs). It is different from the analysis of Sambruna
et al. (1996), we analyzed the relation of luminosity at the each
energy band with $\alpha_{\rm rx}$ and $\nu_p$, respectively.

From our analysis, we found that radio luminosity $L_{\rm r}$ can
be used to distinguish HBLs and LBLs. Our results indicate that
there is  a very strong positive correlation between $L_{\rm r}$
and $\alpha_{\rm rx}$ and a negative correlation between $L_{\rm
r}$ and $\nu_p$, more importantly,  HBLs and LBLs can be
distinguished very well in both $L_{\rm r}$-$\alpha_{\rm rx}$ plot
(see Fig. 1a) and $L_{\rm r}$-$\nu_p$ plot (see Fig. 1b). We have
also analyzed the relations of the optical and X-ray luminosities
with $\alpha_{\rm rx}$ and $\nu_p$, respectively. We found that
the optical luminosity has a good correlation with both
$\alpha_{\rm rx}$ and $\nu_p$ (see Fig. 2a and 2b), but X-ray
luminosity shows no any correlation with $\alpha_{\rm rx}$ or
$\nu_p$ (see Fig. 3a and 3b). Obviously, either $L_{\rm o}$ or
$L_{\rm x}$ cannot be used to distinguish HLBs and LBLs.
 Therefore, $\alpha_{{\rm rx}}$, ${\nu _{p}}$ and $L_{\rm r}$ are equivalent
in the classifications of HBLs and LBLs:  HBLs fall in the region
divided by the lines of $\alpha_{\rm rx}\le 0.75$ (or $\log \nu_p
> 14.7$ Hz) and $\log L_{\rm r}\le 43.25$ erg~s$^{-1}$ while LBLs in the
region divided by the lines of $\alpha_{\rm rx}> 0.75$ (or $\log
\nu_p \le 14.7$ Hz) and $\log L_{\rm r}> 43.25$ erg~s$^{-1}$.

It has been known that the observed spectral energy distributions
of BL Lac objects show a peak between IR and X-rays, and the
possible radiation mechanism is synchrotron radiation. In Fig. 1a
of $L_{\rm r}$ and $\alpha_{\rm rx}$ plot, HBLs distribute in the
region divided by the lines of $\alpha_{\rm rx}\le 0.75$ and $\log
L_{\rm r}\le 43.25$ erg~s$^{-1}$. Using the definition of
$\alpha_{\rm rx}$ and $\alpha_{\rm rx}\le 0.75$, we have $\log
L_{\rm x}\ge\log L_{\rm r} +1$ erg~s$^{-1}$ for HBLs where
$\nu_r=5$ GHz and $\nu_x=1$ keV are used. Since $\log L_{\rm r}\le
43.25$ erg~s$^{-1}$, we have $\log L_{\rm x}\ge 44.25$
erg~s$^{-1}$ for HBLs while $\log L_{\rm x}< 44.25$ erg~s$^{-1}$
for LBLs.  Compared to the result shown in Fig. 3a, the observed
X-ray luminosities of HBLs in our sample are above the lower limit
of $\log L_{\rm x}\ge 44.25$ erg~s$^{-1}$, indicating that X-rays
for HBLs are produced by the synchrotron radiation. From Fig. 3a,
the observed X-ray luminosities for most LBLs do not satisfy the
upper limit of $L_{\rm x}< 44.25$ erg~s$^{-1}$. We believe that
inverse Compton scattering have more important contributions to
X-ray emission from these LBLs.

Finally, we would like to point out that we have made the analysis
about the relation of energy fluxes at radio, optical and X-ray
bands with $\alpha_{\rm rx}$ and $\nu_p$. We find that radio
energy flux can be distinguish HBLs and LBLs very well.

\begin{acknowledgements}
      The Special Funds for Major State Basic Project  of
      China(Grant No.2000077602),{\bf the National 973 project of China
(Grant No.NKBRAFG19990754)}, the Natural Science Foundation of
China and the Natural Science Foundation of Yunnan province  of
China are acknowledged for financial support. We would like to
thank Dr. J.H.Fan for useful discussion.
\end{acknowledgements}

\begin{table*}
\begin{center}
\caption{Sample of X-ray selected BL lacertae objects
(XBLs).}\label{tab:first}
\begin{tabular}{lllllll}
\hline \hline
Name & $z^{*}$ & $\alpha_{\rm{rx}}^{*}$ & $F_{\rm{r}}^{*}$(Jy) & $F_{\rm{o}}^{*}$(mJy) & $%
F_{\rm{x}}^{*}(\mu \rm{Jy})$ & $\log{\nu_{\rm{p}}^\#}$ \\
&  &  & ($5$ GHz) & ($5500$ \AA) & ($1$ keV)
 & (\textrm Hz) \\
\hline \hline
0112.1+0903 & 0.339 & 0.58 & 0.0014 & $0.047\pm0.001$& 0.05 & 15.62 \\
0158.5+0019 & 0.299 & 0.51 & 0.0113 & $0.21\pm0.06$  & 1.2  & 17.34 \\
0205.7+3509 & 0.318 & 0.46 & 0.0036 & $0.10\pm0.005$ & 0.90 & 18.06 \\
0257.9+3429 & 0.247 & 0.65 & 0.010  & $0.25\pm0.02$  & 0.10 & 15.03 \\
0317.0+1834 & 0.190 & 0.58 & 0.017  & $0.36\pm0.09$  & 0.54 & 15.79 \\
0419.3+1943 & 0.512 & 0.53 & 0.008  & 0.09           & 0.75 & 18.31 \\
0607.9+7108 & 0.267 & 0.70 & 0.0182 & 0.09           & 0.07 & 15.23 \\
0737.9+7441 & 0.315 & 0.56 & 0.024  & 0.64           & 1.30 & 16.09 \\
0922.9+7459 & 0.638 & 0.55 & 0.0033 & $0.044\pm0.002$& 0.21 & 17.21 \\
0950.9+4929 & 0.207 & 0.51 & 0.0033 & $0.122\pm0.04$ & 0.27 & 16.13 \\
1019.0+5139 & 0.141 & 0.45 & 0.0024 & 0.22           & 0.93 & 16.69 \\
1207.9+3945 & 0.615 & 0.52 & 0.0058 & 0.10           & 0.55 & 17.38 \\
1221.8+2452 & 0.218 & 0.67 & 0.0264 & $0.42\pm0.09$  & 0.18 & 15.00 \\
1229.2+6430 & 0.164 & 0.56 & 0.042  & $0.55\pm0.17$  & 2.05 & 16.71 \\
1235.4+6315 & 0.297 & 0.55 & 0.007  & $0.14\pm0.02$  & 0.31 & 16.16 \\
1402.3+0416 & 0.200 & 0.57 & 0.0208 & $0.88\pm0.37$  & 0.68 & 15.39 \\
1407.9+5954 & 0.495 & 0.66 & 0.0165 & $0.07\pm0.01$  & 0.10 & 15.76 \\
1443.5+6349 & 0.299 & 0.58 & 0.0116 & 0.06           & 0.35 & 17.71 \\
1458.8+2249 & 0.235 & 0.58 & 0.0298 & $1.01\pm0.20$  & 0.78 & 15.39 \\
1534.8+0148 & 0.312 & 0.61 & 0.034  & $0.15\pm0.05$  & 0.74 & 17.40 \\
1552.1+2020 & 0.222 & 0.54 & 0.0375 & $0.44\pm0.08$  & 2057 & 17.45 \\
1757.7+7034 & 0.407 & 0.50 & 0.0072 & 0.18           & 0.92 & 17.15 \\
2143.3+0704 & 0.237 & 0.61 & 0.050 & $0.32\pm0.04$   & 0.78 & 16.24 \\
\hline
\end{tabular}
\end{center}
\begin{center}
$^{*}$These data are taken from Tables 1 and 2 of Sambruna et al.
(1996).\\
 $^{\#}$These data are taken from Tables 1  of Dong,Mei\& Liang (2002).
\end{center}
\end{table*}
\begin{table*}
\begin{center}
\caption{Sample of radio selected BL lacertae objects
(RBLs).}\label{table:2}
\begin{tabular}{lllllll}
\hline \hline
Name & $z^{*}$ & $\alpha_{\rm{rx}}^{*}$ & $F_{\rm{r}}^{*}$(Jy) & $F_{\rm{o}}^{*}$(mJy) & $%
F_{\rm{x}}^{*}(\mu \rm{Jy})$ & $\log{\nu_{\rm{p}}}^{*}$  \\
&  &  & ($5$ GHz) & ($5500$ \AA) & ($1$ keV) & (Hz)\\
\hline \hline
0048$-$097 & ...      & 0.75 &$1.110\pm0.583$ &$2.41\pm1.63$ & 0.77  & 13.84 \\
0118$-$272 & $>$0.557 & 0.86 & $1.145\pm0.075$& $1.92\pm0.38$& 0.20  & 14.49 \\
0235+164   & 0.940    & 0.76 & $1.81\pm0.54$  & $1.44\pm1.06$& 1.56  & 13.39 \\
0426$-$380 & $>$1.030 & 0.90 & $1.15\pm0.03$  & 0.11         & 0.09  & 13.22 \\
0454+844   & 0.112    & 1.01 & $1.41\pm0.14$  & $0.70\pm0.36$& 0.02  & 13.81 \\
0537$-$441 & 0.896    & 0.82 & $3.93\pm0.17$  & $1.49\pm0.43$& 0.78  & 14.07 \\
0716+714   & $>$0.300 & 0.75 & $0.86\pm0.18$  & 2.96         & 1.17  & 13.79 \\
0735+178   & $>$0.424 & 0.88 & $2.13\pm0.50$  & $3.22\pm1.56$& 0.22  & 14.03 \\
0814+425   & 0.258    & 0.98 & $1.86\pm0.68$  & $0.26\pm0.04$& 0.05  & 13.34 \\
0851+202   & 0.306    & 0.84 & $2.99\pm0.56$  & $6.08\pm5.91$& $0.70\pm0.25$ & 13.72 \\
0954+658   & 0.367    & 0.88 & $0.90\pm0.38$  & $0.86\pm0.24$& 0.16  & 14.09 \\
1144$-$379 & 1.048    & 0.82 & $1.61\pm0.96$  & $0.62\pm0.37$& 0.41  & 13.75 \\
1147+245   & ...      & 0.92 & $0.82\pm0.12$  & $1.53\pm0.36$& 0.05  & 14.58\\
1308+326   & 0.997    & 0.91 & $2.26\pm0.40$  & $2.23\pm1.53$& 0.13  & 13.83 \\
1418+546   & 0.152    & 0.85 & $1.22\pm0.38$  & $2.72\pm0.82$& 0.30  & 13.85 \\
1519$-$273 & ...      & 0.86 & $2.17\pm0.25$  & $0.47\pm0.35$& 0.39  & 13.17\\
1538+149   & 0.605    & 0.93 & $1.53\pm0.42$  & $0.32\pm0.10$& 0.09  & 13.56 \\
1652+398   & 0.033    & 0.67 & $1.27\pm0.10$  &$15.65\pm4.52$& 8.30  & 15.01 \\
1749+096   & 0.320    & 0.92 & $1.44\pm0.36$  & $1.18\pm0.54$&$0.14\pm0.01$ &13.27 \\
1749+701   & 0.770    & 0.81 & $1.11\pm0.35$  & $0.99\pm0.22$& 0.15  & 14.43 \\
1803+784   & 0.679    & 0.88 & $2.79\pm0.30$  & $0.99\pm0.22$&$0.26\pm0.03$ & 13.43 \\
1807+698   & 0.051    & 0.87 & $1.71\pm0.32$  & $7.85\pm2.44$& 0.32  & 14.26 \\
1823+568   & 0.664    & 0.85 & $1.45\pm0.21$  & 0.17         & 0.42  & 13.65 \\
2005$-$489 & 0.071    & 0.71 & $1.21\pm0.02$  & $9.85\pm1.71$&$4.12\pm1.77$ & 14.86 \\
2007+777   & 0.342    & 0.91 & $1.72\pm0.41$  & $1.17\pm0.18$& 0.17  & 13.66 \\
2131$-$021 & 0.557?   & 0.96 & $1.84\pm0.31$  & $0.16\pm0.04$& 0.05  & 13.16 \\
2200+420   & 0.069    & 0.85 & $2.14\pm0.66$  & $8.65\pm4.62$& 0.88  & 14.25 \\
2240$-$260 & 0.774    & 0.89 & 1.03           & $0.26\pm0.10$& 0.07  & 13.32 \\
2254+074   & 0.190    & 0.88 & $0.56\pm0.27$  & $0.60\pm0.19$&
0.09  & 13.25 \\ \hline
\end{tabular}
\end{center}
\begin{center}
$^{*}$These data are taken from Tables 1 and 2 of Sambruna et
al.,(1996).
  \end{center}
\end{table*}

\clearpage
\begin{table*}
\begin{center}
\caption{Linear Regression Analysis Results
($y=Ax+B,N=52$)}\label{table:3}
\begin{tabular}{llcccc}
\hline \hline
$y$ & $x$ & $A$ & $B$& $r$& $P$  \\
\hline \hline
$\log L_r$&$\alpha_{rx}$&$6.64\pm0.49$&$38.24\pm0.37$&$0.886$&$2.26\times10^{-17}$\\
$\log L_r$&$\nu_p$&$-0.67\pm0.06$&$53.13\pm0.92$&$-0.841$&$4.42\times10^{-14}$\\
$\log L_o$&$\alpha_{rx}$&$2.45\pm0.45$&$43.94\pm0.33$&$0.614$&$2.70\times10^{-6}$\\
$\log L_o$&$\nu_p$&$-0.28\pm0.05$&$49.93\pm0.68$&$-0.660$&$2.52\times10^{-7}$\\
$\log L_x$&$\alpha_{rx}$&$-1.41\pm0.43$&$46.08\pm0.32$&$-0.42$&$0.002$\\
$\log L_x$&$\nu_p$&$0.11\pm0.05$&$43.43\pm0.73$&$0.30$&$0.03$\\
 \hline
\end{tabular}
\end{center}
\begin{center}
$r$ denotes the correlation coefficient and $P$ denotes the chance
probability.
\end{center}
\end{table*}

\clearpage
   \begin{figure*}
   \centering
   \includegraphics[width=12cm,height=15cm]{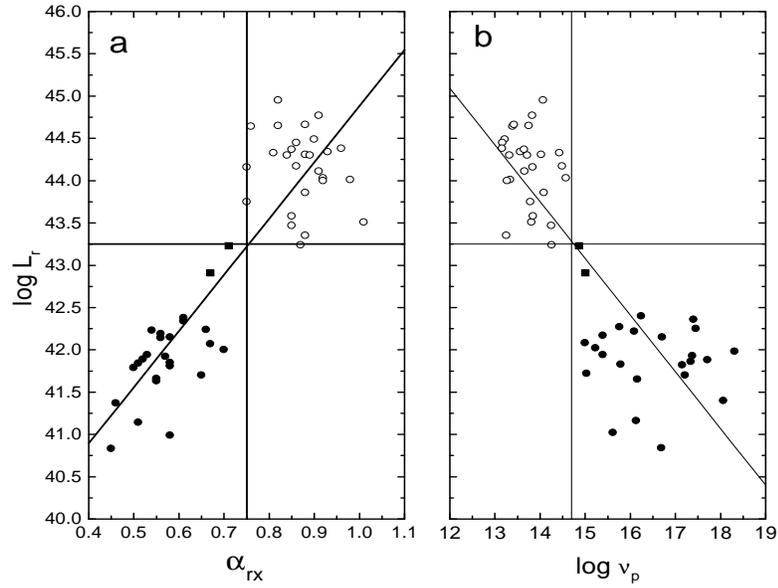}
      \caption{a: Plot of $\log{L_{r}}-\alpha _{rx}$ for RBL and
XBL samples. The empty circles represent RBLs, solid squares
represent 1652+398 and 2005-489, and the solid circles represent
XBLs. The horizontal solid line is $\alpha_{\rm rx}=0.75$ and the
vertical solid line is $\log {L_{r}}=43.25$. The oblique solid
line is the regression line $\log
{L_{r}}=(6.64\pm0.49)\alpha_{rx}+(38.24\pm0.37)$; b: Plot of
$\log{L_{r}}-\log{\nu_p}$ for RBL and XBL samples. The empty
circles represent RBLs, solid squares represent 1652+398 and
2005-489, and the solid circles represent XBLs. The horizontal
solid line is $\log{\nu_p}=14.7$ and the vertical solid line is
$\log {L_{r}}=43.25$. The oblique solid line is the regression
line $\log {L_{r}}=-(0.67\pm0.06)\log{\nu_{p}}+(53.13\pm0.92)$.
              }
   \end{figure*}

\clearpage
   \begin{figure*}
   \centering
   \includegraphics[width=12cm,height=15cm]{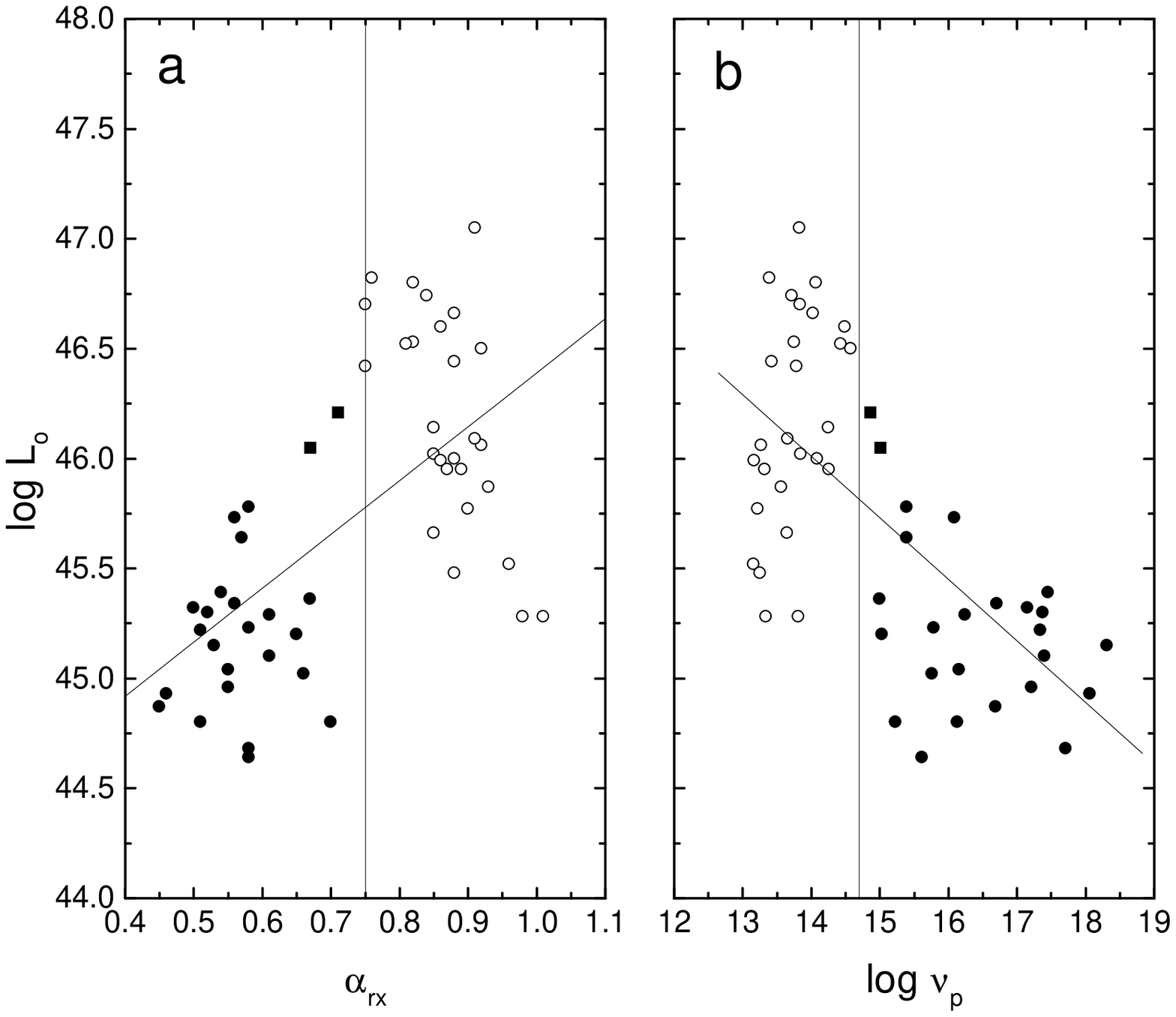}
      \caption{a: Plot of $\log{L_{\rm o}}-\alpha_{\rm rx}$ for
RBL and XBL samples. The empty circles represent RBLs ,solid
squares represent 1652+398 and 2005-489 and the solid circles
represent XBLs. The horizontal solid line is $\alpha_{\rm
rx}=0.75$; b: plot of $\log{L_{\rm o}}-\log{\nu_p}$ for RBL and
XBL samples. The empty circles represent RBLs, solid squares
represent 1652+398 and 2005-489, and the solid circles represent
XBLs. The horizontal solid line is $\log{\nu_p}=14.7$.
              }
   \end{figure*}

\clearpage
   \begin{figure*}
   \centering
   \includegraphics[width=12cm,height=15cm]{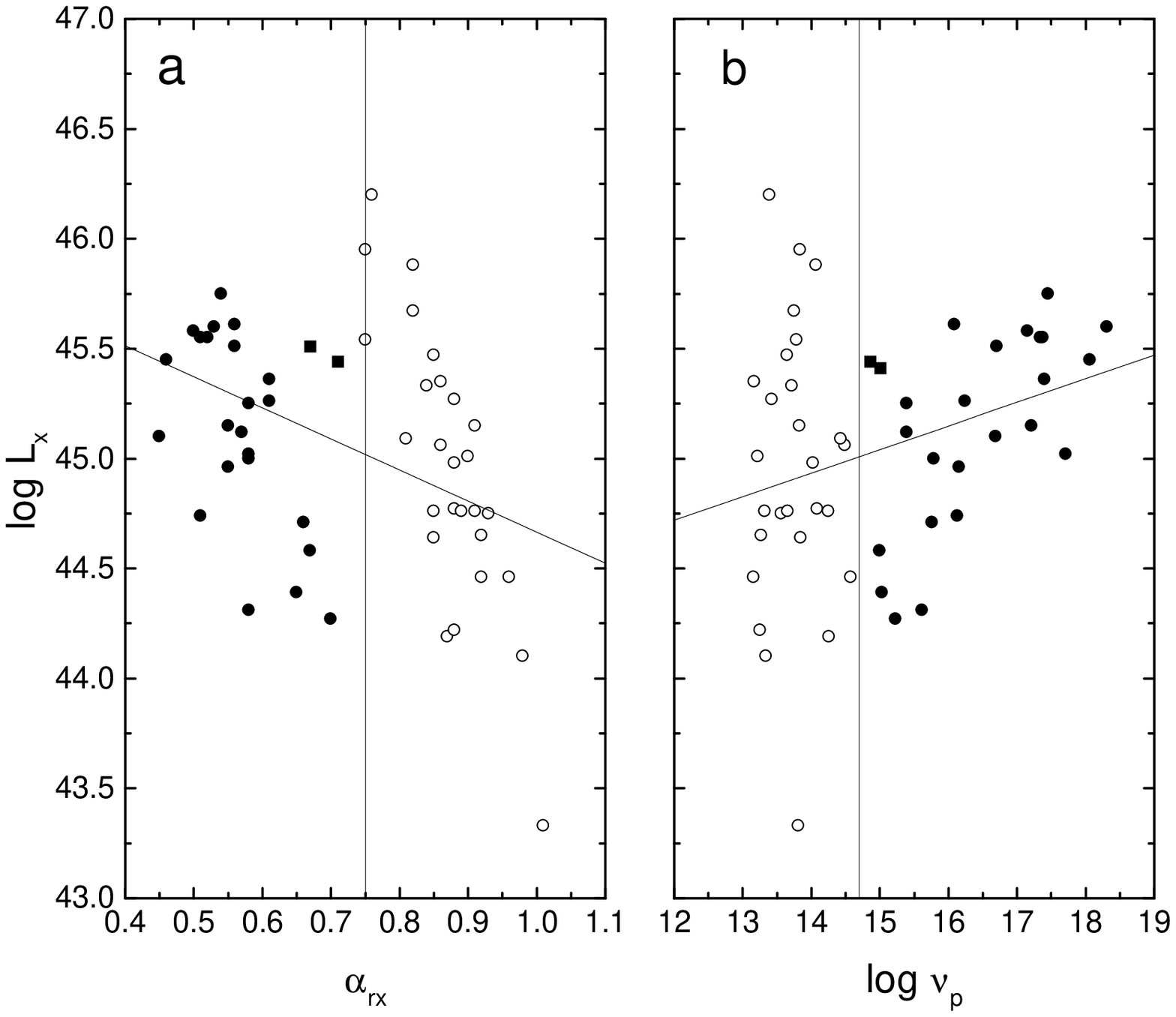}
      \caption{a: plot of $\log{L_{\rm x}}-\alpha_{\rm rx}$ for
RBL and XBL samples. The empty circles represent RBLs, solid
squares represent 1652+398 and 2005-489 and the solid circles
represent XBLs. The horizontal solid line is $\alpha_{\rm
rx}=0.75$; b: plot of $\log{L_{\rm x}}-\log{\nu_p}$ for RBL and
XBL samples. The empty circles represent RBLs, solid squares
represent 1652+398 and 2005-489, and the solid circles represent
XBLs. The horizontal solid line is $\log{\nu_p}=14.7$.
              }
   \end{figure*}


\begin{thebibliography}{}
\bibitem[Dong, Mei, \& Liang, 2002]{Don02}  Dong,Y.M.,Mei,D.C.,\&
Liang,E.W.2002,PASJ,54,{\bf 171}
\bibitem[Falomo et al. 1994]{Fa94}  Falomo,R., Scarpa,R.,\& Bersanelli,M.1994,ApJS,93,125
\bibitem[Fossati et al. 1998]{Fo98}   Fossati, G., et al. 1998,
MNRAS, 299, 433
\bibitem[Ghisellini et al.1986]{Gh86}  Ghisellini,G., et al. 1986,ApJ,310,317
\bibitem[Giommi \& Padovani, 1994]{Gi94} Giommi,P.,\& Padovani,P.1994,MNRAS,268,L51
\bibitem[Giommi et al. 1995]{Gi95}   Giommi, P., Ansari, S. G., \& Micol, A. 1995, A\&AS, 109, 267.
\bibitem[Giommi et al. 1990]{Gi90}   Giommi, P., et al. 1990, ApJ, 356, 432
\bibitem[Lamer et al. 1996]{La96}Lamer, G., Brunner, H., \& Staubert, R. 1996, A\&A, 311, 384
\bibitem[Landau et al. 1986]{La86}   Landau, R., et al. 1986,APJ, 308,78
\bibitem[Ledden, O'Dell, 1985]{Le85}Ledden, J. E., \& O'Dell, S. L. 1985, ApJ, 298, 630
\bibitem[Maraschi et al. (1992)]{Ma92}Maraschi, L., Ghisellini, G., \& Celotti, A. 1992, ApJ, 397, L5
\bibitem[Padovani \& Giommi, 1995]{Pa95}Padovani, P., \& Giommi, P. 1995, ApJ, 444, 567
\bibitem[Padovani \& Giommi, 1996]{Pa96}Padovani, P., \& Giommi, P. 1996, MNRAS, 279, 526
\bibitem[Qin et al. 1999]{Qin99}Qin, Y. P., Xie, G. Z. \& Zheng, X. T. 1999, Ap\&SS,  266, 549
\bibitem[Sambruna et al. 1996]{Sa96}Sambruna, R. M., Maraschi, L., \& Urry, C. M. 1996, ApJ, 463, 444
\bibitem[Schachter et al. 1993]{Sc93}Schachter, J. F.,et al. 1993, ApJ, 412, 541
\bibitem[Stocke et al. 1985]{St85} Stocke, J. T.,et al. 1985, ApJ, 298, 619
\bibitem[Stocke et al. 1991]{St91}   Stocke, J. T., et al. 1991, ApJS, 76, 813
\bibitem[Urry \& Padovani, 1995]{Ur95}  Urry, C. M., \& Padovani, P. 1995, PASP,  107, 803
\end{thebibliography}
\end{document}